\documentclass[11pt]{article}
\usepackage{vietnam,epsfig}

\usepackage{epsfig}
\newcommand{\be}{\begin{equation}}
\newcommand{\ee}{\end{equation}}
\newcommand{\bea}{\begin{eqnarray}}
\newcommand{\eea}{\end{eqnarray}}
\newcommand{\ta}{\tilde\alpha}
\newcommand{\tb}{\tilde\beta}

\newcommand{\Sigb}{{\overline\Sigma}}

\def\be{\begin{equation}}
\def\ee{\end{equation}}
\def\bea{\begin{eqnarray}}
\def\eea{\end{eqnarray}}

\begin{document}
\vspace*{4cm}
\title{MSGUT : from Futility to Precision
\footnote{Talk given at 5th Rencontres de Vietnam,
 Hanoi, Aug. 5-11, 2004}}

\author{ Charanjit S. Aulakh }

\address{Department of Physics,  Panjab University,\\
 Chandigarh, India, 160022 }

\maketitle\abstracts{
 We   computed  the complete  gauge and
chiral superheavy spectra and couplings in
 the Minimal Susy GUT and therefrom evaluated
  the dependence of 1-loop threshold corrections to the
Weinberg angle and Unification scale as functions of   the single
complex parameter $\xi$ that controls MSGUT symmetry breaking.
 The corrections are generically within 10\%
   showing that contrary to longstanding conjectures,{\it{  high
precision calculations are not futile but    necessary and
feasible in the SO(10) MSGUT.}} Effective superpotentials for
$B-L$ violation and  mass formulae of the MSSM matter
supermultiplets
 including neutrino Type I and II seesaws
were derived.}

The MSGUT i.e the  Supersymmetric SO(10) GUT  based on the ${\bf
126}, {\bf {\overline{126}}},
 {\bf 210}$   Higgs multiplets
\cite{aulmoh,ckn}  is now a focus of multifaceted interest
 motivated  by its economy and predictivity. This is the
 most natural and minimal renormalizable
model\cite{lee,abmsv} compatible with the
 observed fermion spectra
and ``minimax'' mixing \cite{babmoh,bsv,moh}.
  In such models the gauge coupling becomes strong
above the the perturbative unification scale $M_X\sim 10^{16} GeV
$  leading ineluctably\cite{trmin,tas} to dynamical symmetry
breaking of the GUT symmetry at a scale $\Lambda_U$ (just above
  $M_X $)  which   is {\it{calculably}}\cite{tas,tas2}
determined by only the low energy data and structural features of
the theory. UV gauge strength leads to
  a novel  picture of
elementarity and dual unification characterized by a new
fundamental length scale $\sim \Lambda_U^{-1}$
 characterizing  the ``hearts of   quarks.''\cite{trmin,tas}.
  We have calculated\cite{msgt2ps} the complete mass spectrum
and couplings of the MSGUT using the techniques developed by us
\cite{alaps} and therefrom the threshold corrections to
$Sin^2\theta_W $ and the perturbative unification scale.
  Contrary
 to longstanding conjectures\cite{dixitsher}
   threshold corrections
   at and to $M_X$ are generically
    small although exceptional parameter regions exist.
     Other recent  calculations of the same
  mass spectra using a different
 \cite{heme} method can be found in \cite{bmsv,fmv2}.
Our spectra coincide (upto
 conventions) with those of the first reference above
 and we comment briefly on differences with the second reference
 at the end. See the main paper\cite{msgt2ps} for detailed comments.

 We study a
renormalizable
 globally supersymmetric $SO(10)$ GUT
 whose chiral supermultiplets  consist of
``AM type''  totally antisymmetric tensors : $
{\bf{210}}(\Phi_{ijkl}),
{\bf{\overline{126}}}({\bf{\Sigb}}_{ijklm}),$
 ${\bf{126}} ({\bf\Sigma}_{ijklm})(i,j=1...10)$
which   break the GUT symmetry
 to the MSSM, together with Fermion mass (FM) Higgs
{\bf{10}}-plet(${\bf{H}}_i$).
  The  ${\bf{\overline{126}}}$ plays a dual or AM-FM
role since  it also enables the generation of realistic charged
fermion      neutrino masses and mixings (via the Type I and/or
Type II mechanisms);
 three  {\bf{16}}-plets ${\bf{\Psi}_A}(A=1,2,3)$  contain the
matter  including the three conjugate neutrinos (${\bar\nu_L^A}$).

 The   superpotential   is
 \bea
  W &=&{1\over{2}}M_{H}H^{2}_{i} + {m \over
{ 4! }} \Phi_{ijkl}\Phi_{ijkl}+{\lambda \over
{{4!}}}\Phi_{ijkl}\Phi_{klmn} \Phi_{mnij}+{M \over { 5!
}}\Sigma_{ijklm}\overline\Sigma_{ijklm}\nonumber\\
 &+&{\eta \over
4!}\Phi_{ijkl}\Sigma_{ijmno}\overline\Sigma_{klmno}+
 {1\over{4!}}H_{i}\Phi_{jklm}(\gamma\Sigma_{ijklm}+
\overline{\gamma}\overline{\Sigma}_{ijklm}) \nonumber \\
  &+& h_{AB}'\psi^{T}_{A}C^{(5)}_{2}\gamma_{i}\psi_{B}H_{i}+ {1\over
5!}f_{AB}' \psi^{T}_{A}C^{(5)}_{2}{\gamma_{i_{1}}}...
{\gamma_{i_5}}\psi_{B}\overline{\Sigma}_{i_1...i_5}
 \label{W}
 \eea
 In
all the MSGUT has exactly 26 non-soft parameters \cite{abmsv}. The
MSSM also has 26 non-soft couplings
 so the 15 parameters of $W_{FM}$ must be
essentially responsible for the 22 parameters describing fermion
masses and mixings in the MSSM.

The GUT scale vevs that break the gauge symmetry down to the SM
symmetry are {\cite{aulmoh,ckn}}
${\langle(15,1,1)\rangle}_{210}:\langle{\phi_{abcd}}\rangle={a\over{2}}
\epsilon_{abcdef}\epsilon_{ef},
\langle(15,1,3)\rangle_{210}~:~\langle\phi_{ab\ta\tb}\rangle=\omega
\epsilon_{ab}\epsilon_{\ta\tb},
  \langle(1,1,1)\rangle_{210}~: ~\langle\phi_{ {\tilde
\alpha}{\tilde \beta} {\tilde \gamma}{\tilde \delta}}
\rangle=p\epsilon_{{\tilde \alpha} {\tilde \beta} {\tilde
\gamma}{\tilde \delta}},
  \langle{\overline\Sigma}_{\hat{1}\hat{3}\hat{5}
\hat{8}\hat{0}}\rangle= \bar\sigma,
\langle{\Sigma}_{\hat{2}\hat{4}\hat{6}\hat{7}\hat{9}} \rangle=
\sigma $. The vanishing of the D-terms of the SO(10) gauge sector
 potential imposes only the condition $
 |\sigma|=|{\overline{\sigma}}| $.
Except for the simpler cases corresponding to enhanced unbroken
symmetry  ($SU(5)\times U(1), SU(5), G_{3,2,2,B-L}, G_{3,2,R,B-L}$
etc)\cite{abmsv,bmsv} this system  of equations is essentially
cubic and can be reduced to  the single  equation \cite{abmsv}
 for a variable $x= -\lambda\omega/m$, in terms of
 which the vevs $a,\omega,p,\sigma,
 {\overline\sigma}$ are specified  :

\be 8 x^3 - 15 x^2 + 14 x -3 = -\xi (1-x)^2 \label{cubic} \ee

where  $\xi ={{ \lambda M}\over {\eta m}} $.  This   exhibits the
crucial importance of the parameter $\xi$.

Using the above vevs and the methods of \cite{alaps} we calculated
the complete   gauge and chiral multiplet GUT scale spectra
{\it{and}} couplings for the 52 different MSSM multiplet sets
falling into 26 different MSSM multiplet types of which 18 are
unmixed while the other 8 types occur in multiple copies and mix
via upto 5 x 5 matrices. ({\it{On a  lighter note}} : the
occurrence yet again of the `mystic' String Theory number 26
 demonstrates that one can do just as
well without string theory !)

If the serendipity of the threefold gauge unification at $M_X^0$
 is to survive closer examination
the  MSGUT  must  answer the query :
 {\it{ Are the one loop values of $Sin^2\theta_W $ and
 $M_X$ generically stable against superheavy threshold
calculations ?}}.
  The parameter $\xi= \lambda M/ \eta m$ is
 the most crucial  determinant of the mass spectrum.
 The formulae for the threshold corrections
 are\cite{weinberg,hall}

\bea
 \Delta^{(th)}(Log_{10}{M_X})  &=& .0217 +.0167 (5 {{\bar b}'}_1 +3{{\bar b}'}_2 -8
  {{\bar b}'}_3) Log_{10}{{M'}\over  {M_X^0}} \label{Deltasw}\\
\Delta^{(th)} (sin^2\theta_W (M_S)) &=&
   .00004 -.00024 (4 {{\bar b}'}_1 -9.6 {{\bar b}'}_2 +5.6
  {{\bar b}'}_3) Log_{10}{{M'}\over  {M_X^0}}
  \label{Deltath} \eea
Where ${\bar b'}_i $ are   1-loop beta function coefficients for
multiplets with mass $M'$.

 We plot these  threshold corrections
 for a range of values of $\xi$ keeping the other ``insensitive"
 parameters   fixed at  randomly chosen
 representative values $ \lambda =0.12 ;
  \eta =0.21  ;  \gamma=0.23  ;
 \bar\gamma=0.35 $.

\begin{figure}[h!]
\begin{center}
\epsfxsize15cm\epsffile{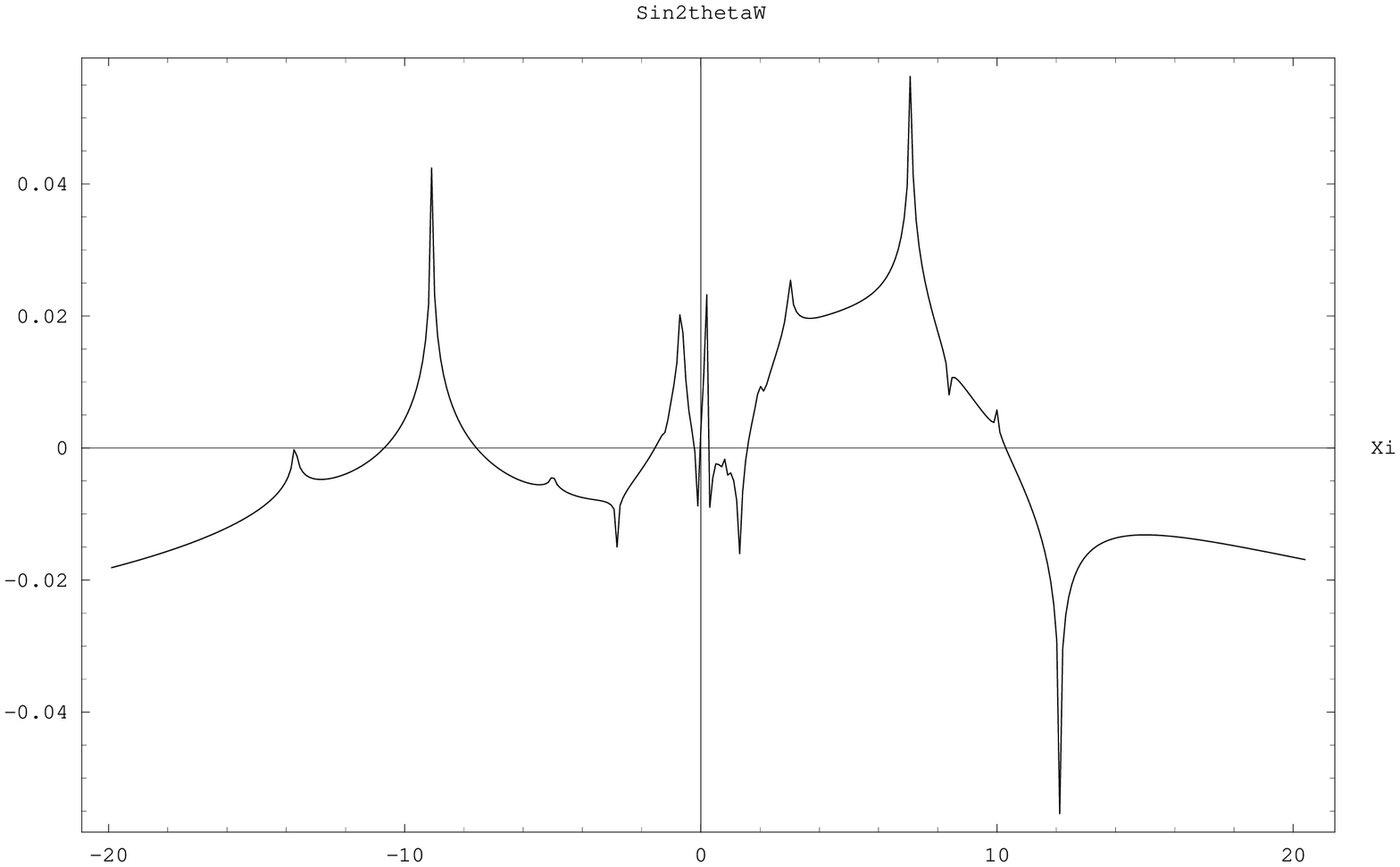}
 \caption{  Plot of the threshold
corrections to $Sin^2\theta_w$ vs $\xi$ for real $\xi$ : real
solution for x.}
\end{center}
\end{figure}

\begin{figure}[h!]
\begin{center}
\epsfxsize15cm\epsffile{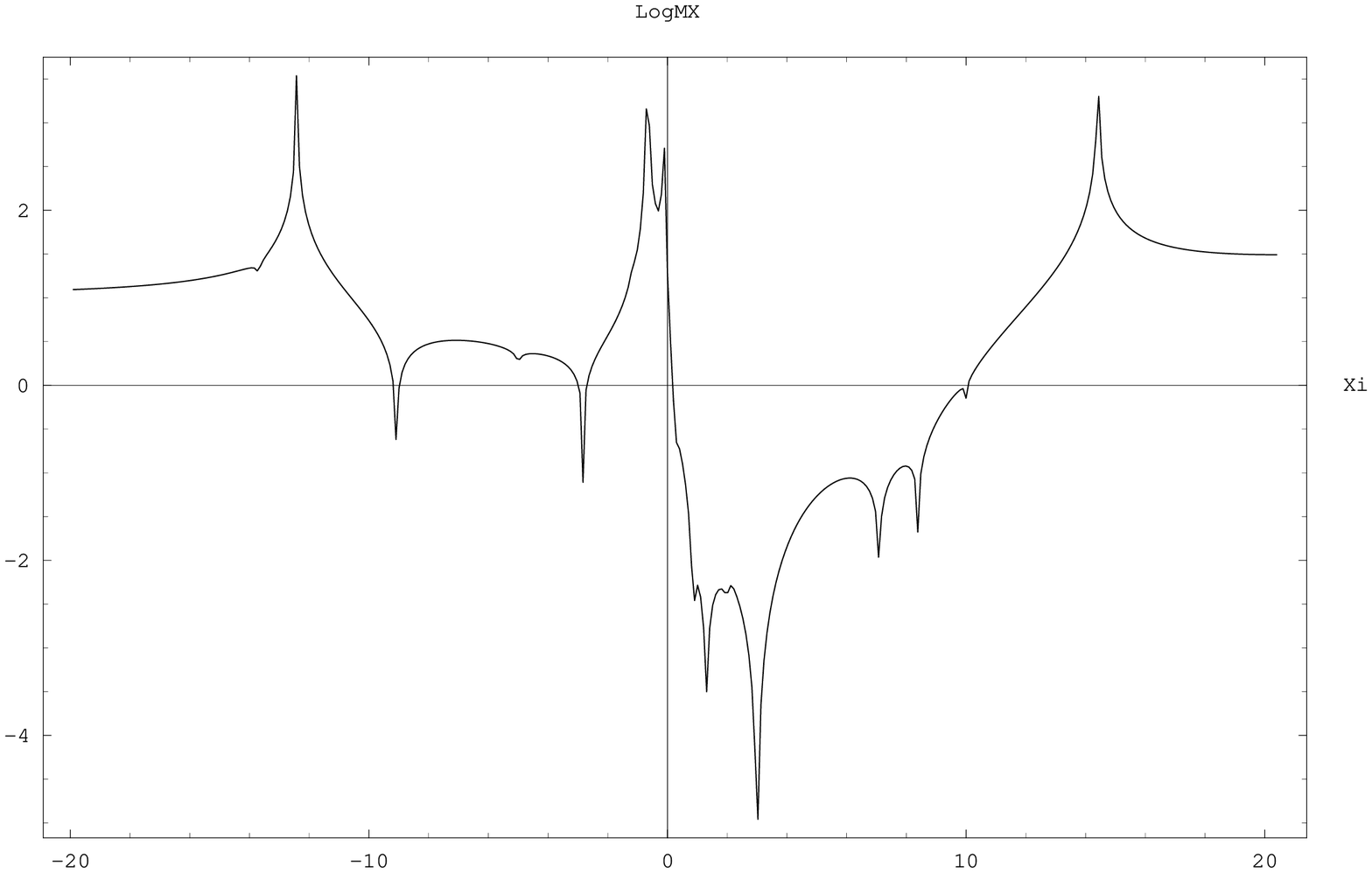}
 \caption{  Plot of the threshold
corrected   $Log_{10} M_X/M_X^0 $ vs $\xi$ for real $\xi$ : real
solution for x.}
\end{center}
\end{figure}

 Generically, effects on
 $sin^2\theta_W (M_S)$ are less
than 10 \% of the 1-loop values  and
the change in $M_X$ is also
 not drastic. Near
 special  points (among which one
 recognizes certain known points
 of enhanced symmetry\cite{abmsv,bmsv})
 the corrections may be
 large but never explosively so.
 Regions where the threshold corrections to
     $Log_{10}M_X$ are   large
need special examination with regard
     to their phenomenological viability and
     consistency with the one scale breaking picture.

 For complex values of $\xi$ as shown
 we find
changes are generically less than 10\% for $Sin^2\theta_w$ while
$M_X$ changes by a factor of 10 or less. Thus our central point is
that:  {\it{contrary to expectations in the
literature\cite{dixitsher},
 precision RG analysis of the SO(10)
MSGUT  is  far from being futile but rather is necessary for
precision unification }}, since the hierarchy of magnitudes
between $O(\alpha^{-1})$  terms
 and 1-loop threshold/2-loop gauge coupling terms
($O(1)$ effects \cite{hall}) is generically maintained.
  However the mechanism that enforces the otherwise
unreasonable insensitivity to  strong growth of the coupling above
$M_X$ must be found\cite{tas2} .

Using  our methods we can compute all couplings of MSSM
submultiplets in terms of GUT couplings and on integrating out the
heavy triplet Higgs supermultiplets$t,{\bar t}$
 one obtains:

 $ W_{eff}^{\Delta
B\neq =0} = L_{ABCD} ({1\over 2}\epsilon Q_A Q_B Q_C L_D)
+R_{ABCD}  (\epsilon {\bar e}_A {\bar u}_B {\bar u}_C {\bar d}_D)
$

 Where, $ L_{ABCD} = {\cal S}_1^{~1} h_{AB} h_{CD} + {\cal
S}_1^{~2} h_{AB} f_{CD} +
 {\cal S}_2^{~1}  f_{AB} h_{CD} + {\cal S}_2^{~2}
   f_{AB} f_{CD}$ and similarly for $ R_{ABCD} $.
   Here ${\cal S}= {\cal T}^{-1} $ and
     $W={\bar t} {\cal T} t +...$

Similarly we can calculate the mass matrices of matter
supermultiplets,  at the GUT scale including their dependence on
the mixing coeffcients defined by the fine tuning condition
necessary to keep one pair of MSSM Higgs doublets light. See the
main paper\cite{msgt2ps}.


  Our
method is different from the computer based method
of\cite{lee,bmsv,fmv2} and is more complete, especially regarding
couplings.  Precision calculations that ignore threshold effects
in SO(10) GUTs seem to be of dubious validity.  The authors of
hep-ph/0405300 claimed that the mass spectra listed above were not
consistent with the requirements of $SU(5)$ or $SU(5)\times U(1)$
symmetry at the special points where $p=a=\pm \omega $.
{\it{However this is entirely incorrect}}. We have checked that in
fact our mass terms naturally respect these symmetries fully at
these special solutions since they organize into appropriate
$SU(5)$  invariants when these conditions hold.

\end{document}